\documentclass[useAMS, usenatbib, onecolumn]{mn2e}
\usepackage{epsfig}
\usepackage{amssymb}
\def\be{\begin{equation}}
\def\ee{\end{equation}}
\def\ba{\begin{eqnarray}}
\def\ea{\end{eqnarray}}
\def\go{\mathrel{\raise.3ex\hbox{$>$}\mkern-14mu
             \lower0.6ex\hbox{$\sim$}}}
\def\lo{\mathrel{\raise.3ex\hbox{$<$}\mkern-14mu
             \lower0.6ex\hbox{$\sim$}}}

\def\tomega{{\tilde{\omega}}}

\begin{document}
\title[Interface Modes in Accretion Discs]{Interface Modes and Their Instabilities in Accretion Disc Boundary Layers}

\author[D. Tsang and D. Lai]
{David Tsang$^{1,2}$\thanks{Email:
dtsang@astro.cornell.edu; dong@astro.cornell.edu} 
and Dong Lai$^{1}$\footnotemark[1] \\ 
$^1$Center for Radiophysics and Space Research, Department of Astronomy,
Cornell University, Ithaca, NY 14853, USA \\
$^2$Department of Physics,
Cornell University, Ithaca, NY 14853, USA \\}

\label{firstpage}
\maketitle

\begin{abstract}
We study global non-axisymmetric oscillation modes trapped near the
inner boundary of an accretion disc. Observations indicate that some
of the quasi-periodic oscillations (QPOs) observed in the luminosities
of accreting compact objects (neutron stars, black holes and white
dwarfs) are produced in the inner-most regions of accretion discs or
boundary layers. Two simple models are considered in this paper: The
magnetosphere-disc model consists of a thin Keplerian disc in contact
with a uniformly rotating magnetosphere with and low plasma density,
while the star-disc model involves a Keplerian disc terminated at the
stellar atomosphere with high density and small density scale height.
We find that the interface modes at the magnetosphere-disc boundary
are generally unstable due to Rayleigh-Taylor and/or Kelvin-Helmholtz
instabilities. However, differential rotation of the disc tends to
suppress Rayleigh-Taylor instability and a sufficiently high disc
sound speed (or temperature) is needed to overcome this suppression
and to attain net mode growth. On the other hand, Kelvin-Helmholtz
instability may be active at low disc sound speeds.  We also find that
the interface modes trapped at the boundary between a thin disc and an
unmagnetized star do not suffer Rayleigh-Taylor or Kelvin-Helmholtz
instability, but can become unstable due to wave leakage to large disc
radii and, for sufficiently steep disc density distributions, due to
wave absorption at the corotation resonance in the disc. The
non-axisymmetric interface modes studied in this paper may be relevant
to the high-frequency QPOs observed in some X-ray binaries and 
in cataclysmic variables.
\end{abstract}

\section{Introduction}

Quasi-periodic variabilities have been observed in the timing data of various types of accreting objects. Several types of quasi-periodic oscillations (QPOs) are observed in X-ray binaries with accreting black holes (BHs) or neutron stars (NSs) (e.g., Remillard \& McClintock 2006; Van der Klis 2006). Oscillations are also seen in the outbursts of accreting white dwarf (WD) systems (e.g., Patterson 1981; see Warner 2004 for a review). 

In accreting NS and BH X-ray binaries the observed QPO frequencies ($40-450$ Hz for the high-frequency QPOs in the BH systems and $\gtrsim 300$ Hz  for kHz QPOs in the NS systems) imply a source close to the central compact object where the Keplerian orbital frequencies are high. Since the BH systems lack a hard surface where oscillations may occur, it is likely that the source of the variability is in the inner regions of the disc itself or in some interface regions between the disc and the plunging flow. Gilfanov et al (2003), however,  found that, based on spectral analysis of the disc emission components, the quasi-periodic variability in Low Mass NS X-ray Binary systems are most likely caused by variations in the disc boundary layer, rather than the disc itself.

In Cataclysmic Variables (CVs) the Dwarf Nova Oscillations (DNOs) seen during during outbursts  have frequencies roughly corresponding to the Keplerian rotation rate at the WD surface (e.g., Patterson, 1981; Warner, 2004; Knigge et al, 1998), which imply an origin at or near the inner disk boundary.

Several models involving accretion disk boundary dynamics have been proposed in different contexts. Popham (1999) studied the effect of a non-axysymmetric bulge at the optically thick to optically thin transition radius as a model for DNOs. Piro \& Bildsten (2004) examined the surface wave oscillations that would occur within the thin equatorial belt around a non-magnetized WD formed by the accretion spreading layer, while Warner \& Woudt (2002) considered accretion onto a slipping belt. In the context of accreting magnetic (neutron) stars, Arons \& Lea (1976) and Elsner \& Lamb (1977) considered the interchange instability at the magnetosphere boundary. Spruit \& Taam (1990) and Spruit, Stehle \& Papaloizou (1995) investigated the stability of thin rotating magnetized discs. Of particular relevance to the present paper is the work of Li \& Narayan (2004), who examined a simplified cylindrical model of the Rayleigh-Taylor and Kelvin-Helmholtz instabilities at the boundary between a magnetosphere and an incompressible rotating flow. There have also been a number of numerical simulations of the interface at the magnetosphere-disc boundary (see Romanova et al., 2008 and Kulkarni \& Romanova, 2008 and references therein).

In this paper we study global non-axisymmetric oscillation modes confined near inner boundary of the accretion disc (interface modes). We consider two simple models. The first model involves the magnetosphere-disc boundary similar to the model of Li \& Narayan (2004): we consider an uniformly rotating incompressible magnetosphere with low gas density (where magnetic pressure dominates), which truncates a thin barotropic accretion disc (where gas pressure dominates). This situation may arise from magnetic field build up due to accretion (e.g., Bisnovatyi-Kogan \& Ruzmaikin, 1974, 1976; Igumenshchev et al., 2003; Rothstein \& Lovelace, 2008) or by the magnetosphere of a central (neutron) star. Unlike Li \& Narayan (2004), who restricted their model to incompressible fluid, our discs are compressible and we show that because of the differential rotation of the disc, finite disc sound speed plays an important role in the development of the instability of the interface modes.

In our second model we examine the interface modes for accretion onto a non-magnetic stellar surface. 
Though the structure of the boundary layer is non-trivial and may affect boundary modes (see, e.g., Carroll et al., 1985; Collins et al., 2000), we consider the instabilities for a thin disc truncated by a sharp transition to a dense uniformly rotating stellar atmosphere. This simplified model may provide insight for modes with characteristic radial length scale much greater than the radial length scale of the boundary layer.

In Section 2 we describe the basic setup  for the magnetospheric boundary model, and in Section 3 we discuss the resulting interface mode instabilities. We describe the star-disc boundary and analyze its possible instabilities in Section 4. We then conclude in Section 5 with a discussion of possible applications of our findings.

\section{Magnetosphere-Disc Setup}

We begin by considering a simplified model of the magnetosphere-disc boundary similar to the one considered by Li \& Narayan (2004). 
The magnetic field is assumed to be negligible in the disc region ($r > r_{\rm in}$), while the magnetosphere region ($r < r_{\rm in}$) is assumed to be incompressible and have low density compared to the disc region, with purely vertical magnetic field. Unlike Li \& Narayan (2004), who assumed infinite sound speed in the disc, our disc has sound speed $c_s$ much less than the disk rotation speed $r\Omega$. 

In terms of the vertically integrated density ($\Sigma$), pressure ($P$), magnetic field (${\bf B}$) and fluid velocity (${\bf u}$) the ideal  MHD equations are:
\ba
\frac{\partial \Sigma}{\partial t} + \nabla \cdot (\Sigma {\bf u}) &=& 0 \label{continuity} \\
\frac{\partial {\bf u}}{\partial t} + ({\bf u} \cdot \nabla) {\bf u} &=& -\frac{1}{\Sigma} \nabla \Pi - \nabla \Phi + \frac{1}{\Sigma} {\bf T}\label{momentumcon}\\
\frac{\partial {\bf B}}{\partial t} &=& \nabla \times ({\bf u} \times {\bf B}).
\ea
where $\Pi \equiv P + B^2/8\pi$ is the total pressure, ${\bf T} = \frac{1}{4\pi}({\bf B} \cdot \nabla) {\bf B}$ is the magnetic tension, and $\Phi$ is the gravitational potential due to the central object (e.g. Fu \& Lai, 2008). Using cylindrical coordinates ($r, \phi, z$), we consider the case where the magnetic field is purely poloidal and ${\bf B} = B_z {\hat z}$ in the disc plane, which gives ${\bf T} = 0$. We assume an axisymmetric background flow with fluid velocity ${\bf u} = r\Omega(r) \hat{\phi}$. The unperturbed flow satisfies the condition
\be
g_{\rm eff} \equiv -\frac{1}{\Sigma} \frac{d\Pi}{dr} = \frac{d\Phi}{dr} - \Omega^2 r~. \label{eq4}
\ee

The linearized equations of (\ref{continuity}) and (\ref{momentumcon}) with perturbations of the form $e^{im\phi - i\omega t}$ (assuming no vertical dependence) take the form: 
\ba
-i\tomega \delta \Sigma + \frac{1}{r} \frac{\partial}{\partial r} (\Sigma r \delta u_r) + \frac{i m \Sigma}{r} \delta u_\phi &=& 0 \label{perturbcons}~,\\
-i\tomega \delta u_r - 2\Omega \delta u_\phi &=& -g_{\rm eff} \frac{\delta \Sigma}{\Sigma} - \frac{1}{\Sigma} \frac{\partial}{\partial r} \delta \Pi \label{perturbmom1}~,\\
-i\tomega \delta u_\phi + \frac{\kappa^2}{2\Omega}\delta u_r &=& - \frac{im}{\Sigma r}\delta \Pi\label{perturbmom2}~,
\ea
where $\tomega = \omega - m\Omega$ is the Doppler shifted frequency, $\kappa = \left[\frac{2\Omega}{r}\frac{d}{dr}(r^2\Omega)\right]^{1/2}$ is the radial epicyclic frequency, and $\delta \Sigma$, $\delta \Pi$ and  $\delta {\bf u}$ are the Eulerian perturbations of the fluid variables. Additionally, assuming a barotropic flow we have
\be
\delta \Sigma = \frac{1}{c_s^2}\delta P = \frac{1}{c_s^2}(\delta \Pi - \frac{1}{4\pi}{\bf B}\cdot \delta {\bf B})~.
\ee
with the sound speed $c_s \equiv (dP/d\Sigma)^{1/2}$.

\subsection{The Magnetosphere}
In the inner, magnetically dominated region ($r < r_{\rm in}$), we assume the flow to be incompressible, and have uniform rotation ($\Omega = \Omega_- =$ const) and uniform surface density ($\Sigma = \Sigma_- =$ const). Equations (\ref{perturbcons}) - (\ref{perturbmom2}) then reduce to
\ba
\frac{1}{r} \frac{\partial}{\partial r} (\Sigma r \delta u_r) + \frac{i m \Sigma}{r} \delta u_\phi &=& 0 \label{inner1} \\
-i\tomega \delta u_r - 2\Omega \delta u_\phi &=& - \frac{1}{\Sigma} \frac{\partial}{\partial r} \delta \Pi\label{inner2} \\
-i\tomega \delta u_\phi + \frac{\kappa^2}{2\Omega}\delta u_r &=& - \frac{im}{\Sigma r}\delta \Pi \label{inner3},
\ea
As in Li \& Narayan (2004) we define $W \equiv r \delta u_r$  and find $\delta u_\phi = (i/m) dW/dr$ and 
\be
\frac{1}{r}\frac{d}{dr}\left(r \frac{dW}{dr}\right) - \frac{m^2}{r^2}\left[ 1 - \frac{r}{m\tomega} \frac{d}{dr}\left(\frac{\kappa^2}{2\Omega}\right)\right]W = 0~.\label{Weq}
\ee
For uniform rotation, $\kappa = 2 \Omega$, equation (\ref{Weq}) has the solution $W \propto r^{\pm m}$. 
Since $r < r_{\rm in}$, we take the positive sign to be the physical solution so that the perturbation falls off away from the interface. Thus the exact solution for the $r < r_{\rm in}$ region is
\be
\delta u_r =  \delta u_r(r_{\rm in})\left(\frac{r}{r_{\rm in}}\right)^{m-1}. \label{innerdeltau}
\ee

\begin{figure}
\centering
\epsfig{file=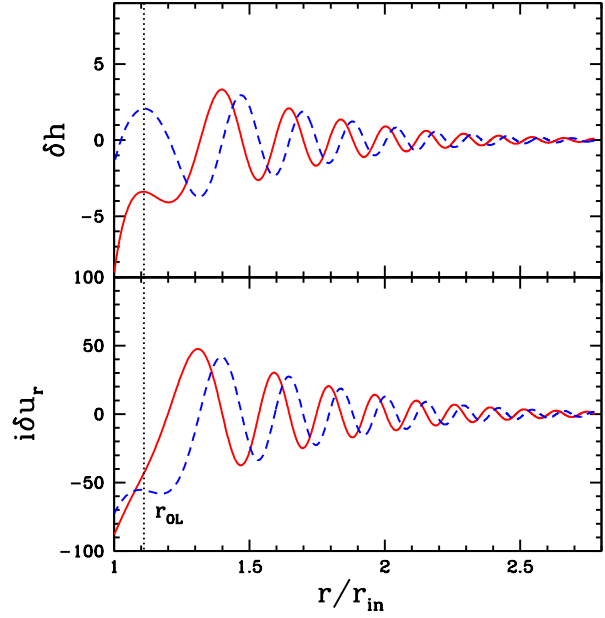,width=0.5\linewidth,clip=}
\caption{The wavefunctions for the interface mode for $m=4$, $c_s = 0.1 r\Omega$, $\Sigma_-/\Sigma_+ = 1/99$ and $\Omega_-/\Omega_{\rm in} = 1$, with the mode frequency $\omega/\Omega_{\rm in} = 4.275 + 0.1914i$, where $\Omega_{\rm in} \equiv \Omega(r_{\rm in+}) \simeq \Omega_k(r_{\rm in})$. The real components are shown in solid lines, while the imaginary components are dashed lines. Note that for the interface modes only the outer Lindblad resonance ($r_{\rm OL}$, denoted by the dotted line) exists outside $r_{\rm in}$}
\end{figure}

\subsection{The Disc}
In the disc ($r > r_{\rm in}$), we take the magnetic field to be small, such that $P \gg B^2/(8\pi)$, and the angular velocity of the unperturbed flow to be nearly Keplerian, such that $\Omega(r) \approx \Omega_k(r) \equiv  \sqrt{\frac{1}{r}\frac{d\Phi}{dr}}$. Rewriting equations  (\ref{perturbcons}) - (\ref{perturbmom2}) we have
\ba
-i\tomega \frac{\Sigma}{c_s^2} \delta h + \frac{1}{r} \frac{\partial}{\partial r} (\Sigma r \delta u_r) + \frac{i m \Sigma}{r} \delta u_\phi &=& 0~,\\
-i\tomega \delta u_r - 2\Omega \delta u_\phi &=& -\frac{\partial}{\partial r} \delta h\label{pert1}~,\\
-i\tomega \delta u_\phi + \frac{\kappa^2}{2\Omega}\delta u_r &=& - \frac{im}{r}\delta h~\label{pert2},
\ea
where 
\be
\delta h \equiv c_s^2 \frac{\delta \Sigma}{\Sigma} = \frac{\delta P}{\Sigma}
\ee
is the enthalpy perturbation. Eliminating the velocity perturbations in favor of the enthalpy, we obtain the second order ODE for the enthalpy perturbation in the disc,
\be
\left[\frac{d^2}{dr^2} -
\frac{d}{dr}\left(\ln\frac{D}{r\Sigma}\right)\frac{d}{dr} -
\frac{2m\Omega}{r\tomega}\left(\frac{d}{dr}\ln\frac{\Omega\Sigma}{D}\right)
- \frac{m^2}{r^2} - \frac{D}{c_s^2}\right]\delta h = 0~.
\label{perturbeq1}
\ee
where $D \equiv \kappa^2 - \tomega^2$. For concreteness we will assume a power-law disc surface density profile $\Sigma \propto r^{-p}$.

\subsection{Matching Conditions Across the Interface}
The matching conditions across the interface at $r_{\rm in}$ between the magnetosphere and the disc region are given by demanding the continuity of the Lagrangian displacement in the radial direction
$\xi_r = i\delta u_r/\tomega$, and the total Lagrangian pressure perturbation $\Delta \Pi = \delta \Pi + \xi_r \frac{d\Pi}{dr}$ across the boundary. The former gives
\be
\frac{i\delta u_{r+}}{\tomega_+} = \frac{i\delta u_{r-}}{\tomega_-}\label{lag}~,
\ee
where the subscript ``$\pm$'' implifes that the quantities are evaluated at $r = r_{{\rm in}\pm}$. The total Lagrangian pressure perturbation for $r = r_{\rm in-}$ is given by
\ba
\Delta \Pi_- &=& \Sigma_-  \left[ \left(\frac{i \kappa^2}{2m\Omega} + \frac{i}{r \tomega}\frac{d\Pi}{dr}\right)W + \frac{ir\tomega}{m^2} \frac{dW}{dr}\right]_{r_{\rm in-}}\nonumber\\
&=& \Sigma_- \left[ \frac{2r \Omega\tomega}{m} - g_{\rm eff-} + \frac{r \tomega^2}{m}\right]\frac{i \delta u_{r}}{\tomega}\bigg|_{r_{\rm in-}}\label{plagminus}.
\ea
In the disc region, we have
\ba
\Delta \Pi_+ &=& \Delta P_+
= \Sigma_+ \left( \delta h + \frac{i \delta u_{r}}{\tomega} \frac{1}{\Sigma}\frac{dP}{dr}\right)_{r_{\rm in+}}\nonumber\\
&=& \Sigma_+ \left( \frac{\tomega \delta h}{i\delta u_{r}} - \frac{p c_s^2}{r}\right) \frac{i\delta u_{r}}{\tomega_+}\bigg|_{r_{\rm in+}} \label{plagplus}.
\ea
The condition $\Delta \Pi_+ = \Delta \Pi_-$ then gives
\be
\Sigma_+ \left(\frac{\tomega \delta h}{i\delta u_{r}} - \frac{pc_s^2}{r}\right)_{r_{\rm in+}} = \Sigma_- \left(\frac{2r\Omega\tomega}{m} - g_{\rm eff} + \frac{r\tomega^2}{m} \right)_{r_{\rm in-}} \label{eq29}.
\ee



\section{Interface Modes at the Magnetosphere-Disc Boundary}
\begin{figure}
\centering
\begin{tabular}{ccc}
\epsfig{file=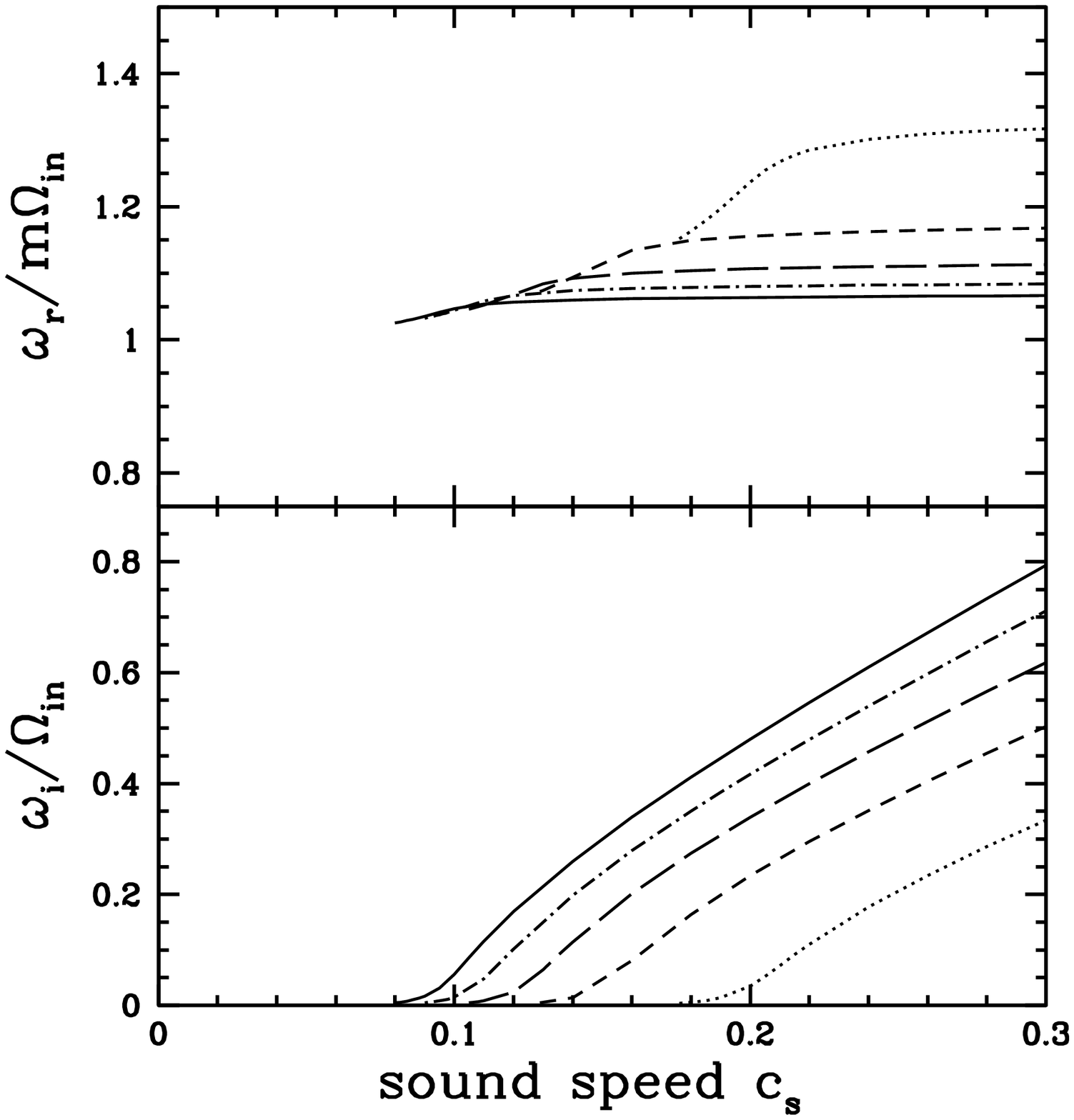, width=0.5\linewidth,clip=} &
\epsfig{file=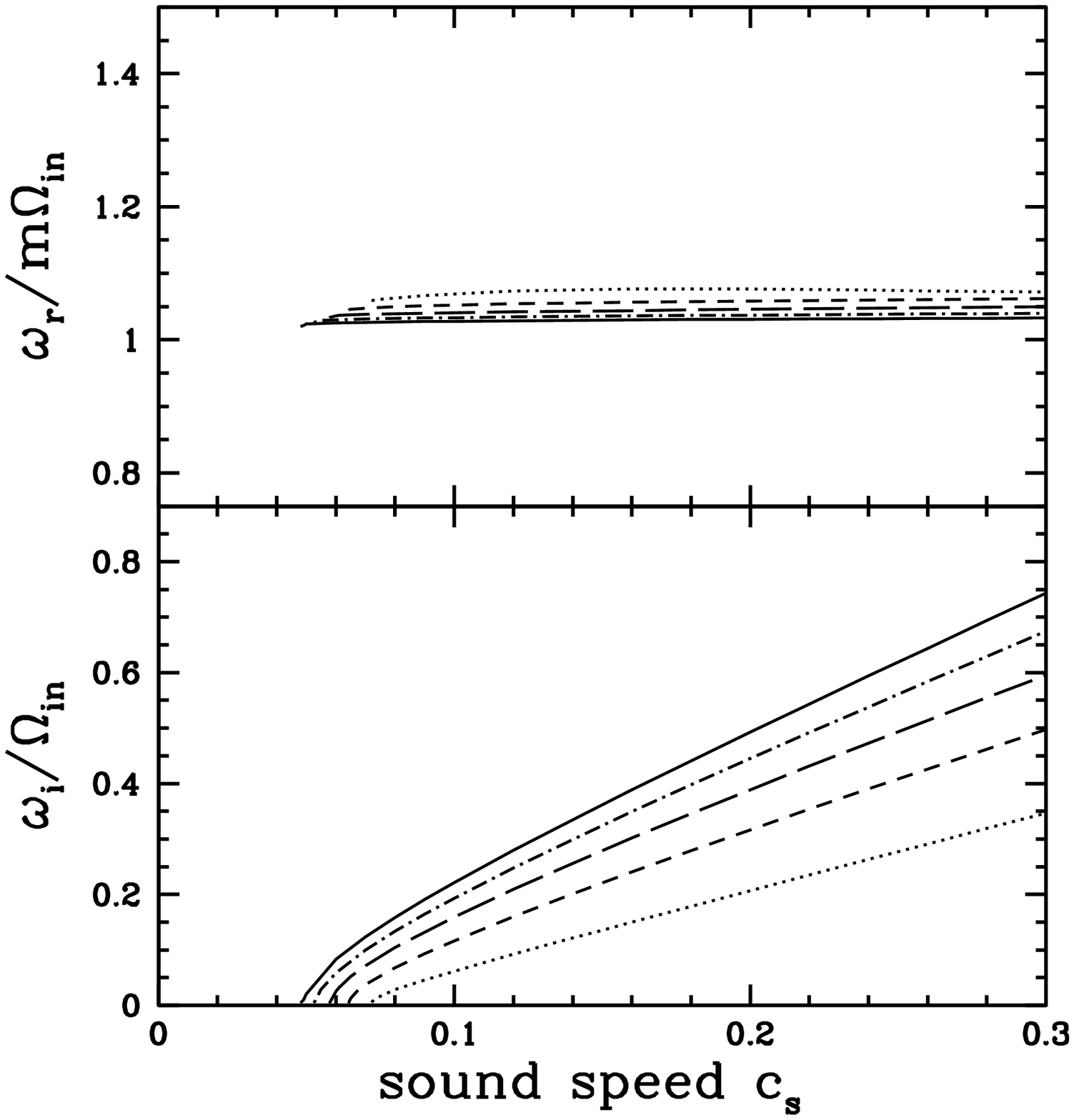, width=0.5\linewidth,clip=}
\end{tabular}
\caption{Real and imaginary frequencies for interface modes for various $m$ as a function of the disc sound speed $c_s$. The left panels are for $\Sigma_-/\Sigma_+ = 0$, while the right panels are for $\Sigma_-/\Sigma_+ = 1/9$, both with $\Omega_-/\Omega_{\rm in} = 1$. The solid lines show the eigenfrequencies for $m=5$ modes, the dash-dotted lines for $m=4$, the long-dashed lines for $m=3$, the short-dashed lines for $m=2$, and the dotted lines for $m=1$.}
\end{figure}

\begin{figure}
\centering
\begin{tabular}{ccc}
\epsfig{file=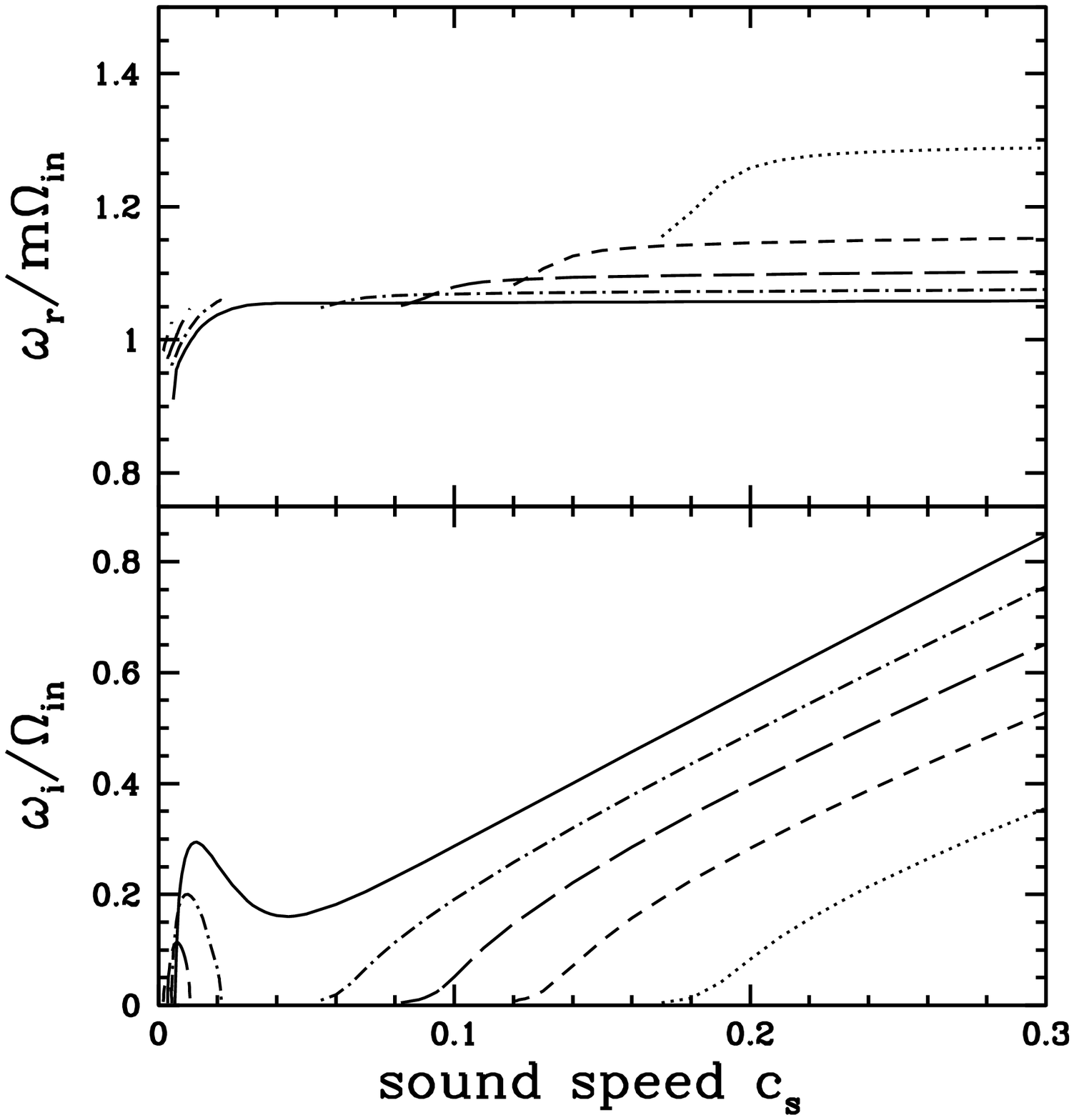,width=0.5\linewidth,clip=} &
\epsfig{file=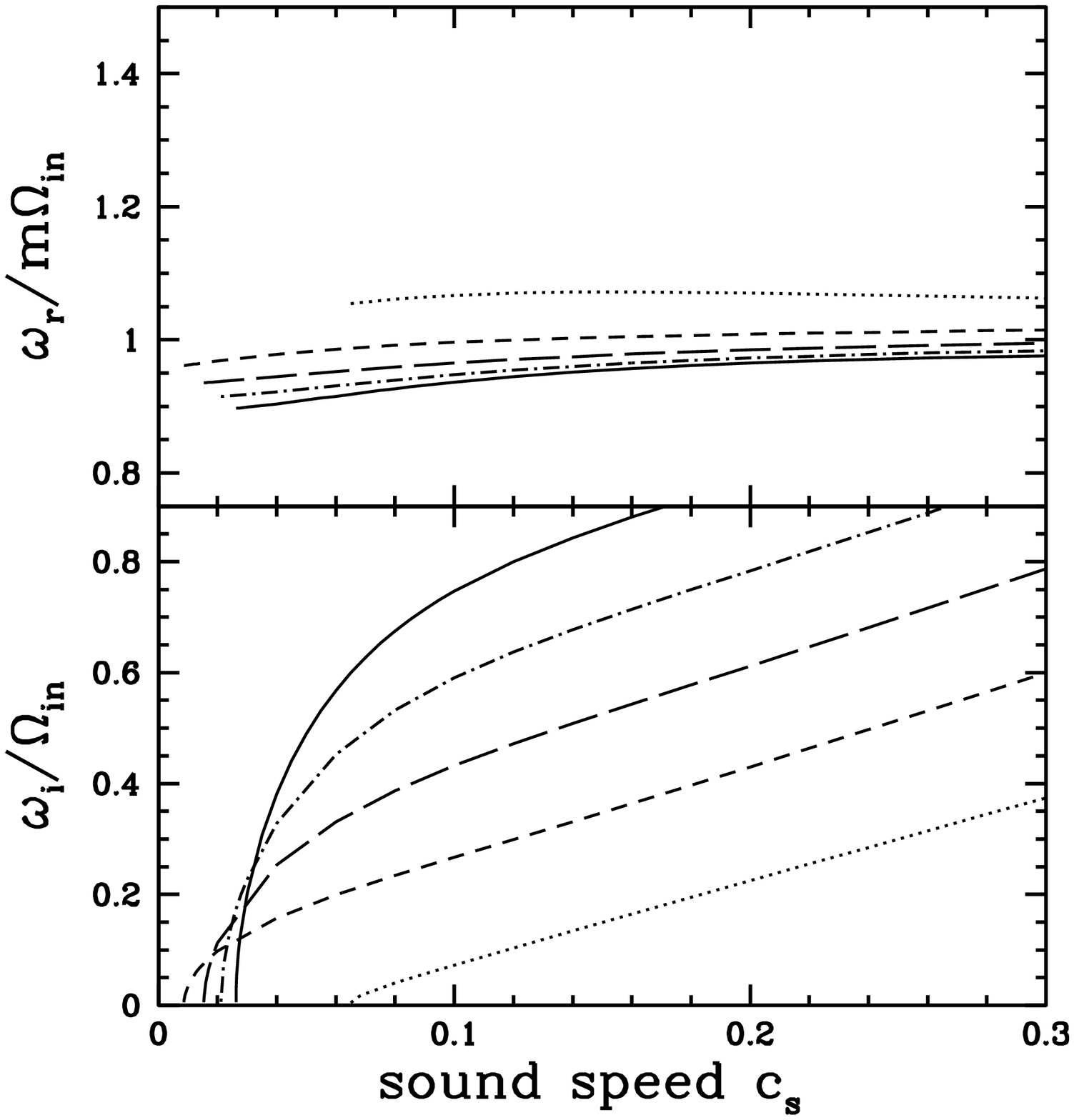,width=0.5\linewidth,clip=}
\end{tabular}
\caption{Same as Figure 2, except that the left panels are for $\Sigma_-/\Sigma_+ = 1/99$, while the right panels are for $\Sigma_-/\Sigma_+ = 1/9$, both with $\Omega_-/\Omega_{\rm in} = 0.5.$}
\end{figure}

Perturbations mainly confined to the magnetosphere-disc interface can become unstable due to Rayleigh-Taylor or Kelvin Helmholtz instability. In order to calculate the growth rates, we must solve the eigenvalue problem given by equation (\ref{perturbeq1}) with an outgoing wave boundary condition at some outer radius, and equation (\ref{eq29}) at the interface radius $r_{\rm in}$. 

\subsection{Numerical Solution}

We adopt the radiative outer boundary condition in the outer wave zone of the disc, such that far from the outer Lindblad resonance radius $r_{\rm OL}$ (where $\omega-m\Omega = \kappa$) we have the solution of the form:
\be
\delta h \propto A \exp\left( i \int^r k dr\right), 
\ee
with $A = (D/r\Sigma k)^{1/2}$ and $k = (-D/c_s^2)^{1/2}$ (see Tsang \& Lai 2008, Lai \& Tsang, 2008). This gives the boundary condition at $r = r_{\rm out} > r_{\rm OL}$:
\be
\delta h'(r_{\rm out}) = \delta h(r_{\rm out}) \left( ik + \frac{1}{A}\frac{dA}{dr}\right)_{r_{\rm out}}. \label{radout}
\ee
We adopt (\ref{eq29}) as the inner boundary condition for the disc and solve the eigenvalue problem using a standard shooting method (Press et al 1998). For the numerical solutions below, the density profile of the disk was assumed to be $\Sigma \sim r^{-3/2}$ so that corotation absorption plays no role in determining the mode stability (Tsang \& Lai 2008). An example wavefunction for an interface mode is shown in Figure 1, for typical disc parameters. 

The numerical eigenvalues are shown in Figure 2 and Figure 3 for various disc and magnetosphere parameters, for $m = 1,2,\ldots,5$.

\subsection{Discussion of Numerical Results}
Figure 2 shows the complex eigenvalues as a function of sound speed $c_s$, for density contrasts corresponding to $\Sigma_- = 0$ and $\Sigma_- = \frac{1}{9}\Sigma_+$, with magnetosphere rotation rate equal to the Kepler frequency at the interface [$\Omega_- = \Omega(r_{\rm in+}) \simeq \Omega_k(r_{\rm in})$].  
For this case we see that there exists a cutoff in the disc sound speed below which no growing interface modes are found. This arises from the stabilizing effect of the background differential rotation, and can be understood as follows. 

Setting $\Sigma_- = 0$ and rewriting (\ref{eq29}) in terms of the radial velocity perturbation $\delta u_r$, we have:
\be
\zeta_+ \tomega + \frac{\tomega^2}{m\Sigma_+ \delta u_r}\frac{d}{dr}\left(\Sigma_+ r \delta u_r \right) + g_{\rm eff+} \left(\frac{r\tomega^2}{mc_s^2} - \frac{m}{r}\right) = 0 \label{eq34}
\ee
where $g_{\rm eff+} = pc_s^2/r$ [see Eq. (\ref{eq4})] and $\zeta \equiv \kappa^2/(2\Omega)$ is the vorticity, and where all quantities are evaluated at the interface $r_{\rm in+}$.  For the wave frequencies of interest, the waves are evanescent in the region of the disk just outside the interface. Let $\tilde{k} \equiv -\delta u_r'/\delta u_r > 0$. Equation (\ref{eq34}) can be solved  in terms of $\tilde{k}$, giving
\be
\tomega = \frac{m \zeta}{2\gamma} \pm i \sqrt{\frac{g_{\rm eff+} m^2}{r\gamma} - \frac{m^2\zeta^2}{4\gamma^2}}\label{simpeq}
\ee
where $\gamma \equiv \tilde{k} r  -1$. The terms inside the square root correspond to the mode growth due to Rayleigh-Taylor instability and the suppression due to vorticity, respectively. With $g_{\rm eff+} = pc^2/r$ we find the critical sound speed 
\be
c_{\rm crit} \approx \sqrt{\frac{\zeta^2 r^2}{4p\gamma}}\label{eq30},
\ee
above which the perturbations will be unstable. 




Figure 3 shows cases where the inner region is uniformly rotating at an angular frequency of one half the Kepler frequency at the interface [$\Omega_- = 0.5 \Omega(r_{\rm in+})$]. When $\Sigma_->0$ this leads to the development of the Kelvin-Helmholtz instability, and both this and the Rayleigh-Taylor instability play a role in the mode growth. In the Appendix we derive the expression for the plane-parallel Rayleigh-Taylor and Kelvin-Helmholtz instabilities for a compressible upper region (with density $\rho_+$ and horizontal velocity $u_+$), and incompressible lower region (with density $\rho_-$ and horizontal velocity $u_-$). 
For $\rho_- \ll \rho_+$ we have $\omega \approx ku_+ \pm i \omega_i$ where $k$ is the horizontal wavenumber and 
\be
\omega_i = \sqrt{k\tilde{k}(u_+ - u_-)^2\left(\frac{\rho_-}{\rho_+}\right) + g \tilde{k}} \equiv \sqrt{\omega_{\rm KH}^2 + \omega_{\rm RT}^2}\label{eq31}.
\ee
Here $\tilde{k} \approx \frac{1}{2H_z}\left[\sqrt{1 + H_z^2 k^2}  - 1\right]$, g is the acceleration due to gravity in the vertical direction, and $H_z$ is the vertical scale height in the upper region.The Kelvin-Helmholtz term is approximately
\be
\omega_{\rm KH}^2 \approx k \tilde{k} (u_+ - u_-)^2 \left(\frac{\rho_-}{\rho_+}\right).
\ee
For $kH_z \gg 1$ this reduces to the incompressible limit with $\omega_{\rm KH}^2 \approx k^2 (u_+ - u_-)^2 \rho_-/\rho_+$.  For $kH_z \ll 1$ we have $\omega_{\rm KH}^2 \approx (H_zk) k^2 (u_+ - u_-)^2 \rho_-/\rho_+$, a factor of $H_zk$ smaller than the incompressible result. 

For the rotating system under consideration, the imaginary part of the mode frequency can be written schematically as [cf. equation (\ref{eq31})]
\be
\omega_i \approx \sqrt{\omega_{\rm KH}^2 + \omega_{\rm RT}^2 + \omega_{\rm vort}^2}.
\ee 
We also have $H_z \sim c_s^2/g_{\rm eff} \sim r$ and $k \sim m/r$ so $kH_z \sim m$. Thus $\omega_{\rm KH}^2 \sim (\Delta \Omega)^2 \Sigma_-/\Sigma_+$, and $\omega_{\rm KH}^2$ depends weakly on sound speed. On the other hand, from equation (\ref{simpeq})  we see that the vorticity suppresses mode growth through the term  $\omega_{\rm vort}^2 = -m^2\zeta^2/(4 \gamma^2)$. For sufficiently small $c_s$, equation (\ref{perturbeq1}) indicates $\delta h \propto e^{-\kappa r/c_s}$, i.e. $\tilde{k} \sim \kappa/c_s$. Therefore the vorticity term scales with sound speed as $\omega_{\rm vort}^2 \sim -m^2 c_s^2/r^2$, and can be dominated by the Kelvin-Helmholtz term for small enough sound speed. In the left panel of Figure 3, the mode growth ($\omega_i > 0$) for small $c_s$ is mainly driven by the Kelvin-Helmholtz instability. For $m \geq 5$ the sound speed ranges where $\omega_{\rm RT}^2$ and $\omega_{\rm KH}^2$ dominate over $\omega_{\rm vort}^2$ overlap, and hence the critical sound speed in equation (\ref{eq30}) is not relevant. For larger values of $\Sigma_-$ these regions can overlap for all $m$.

\subsection{Effect of a Relativistic Potential}
\begin{figure}
\centering
\begin{tabular}{ccc}
\epsfig{file=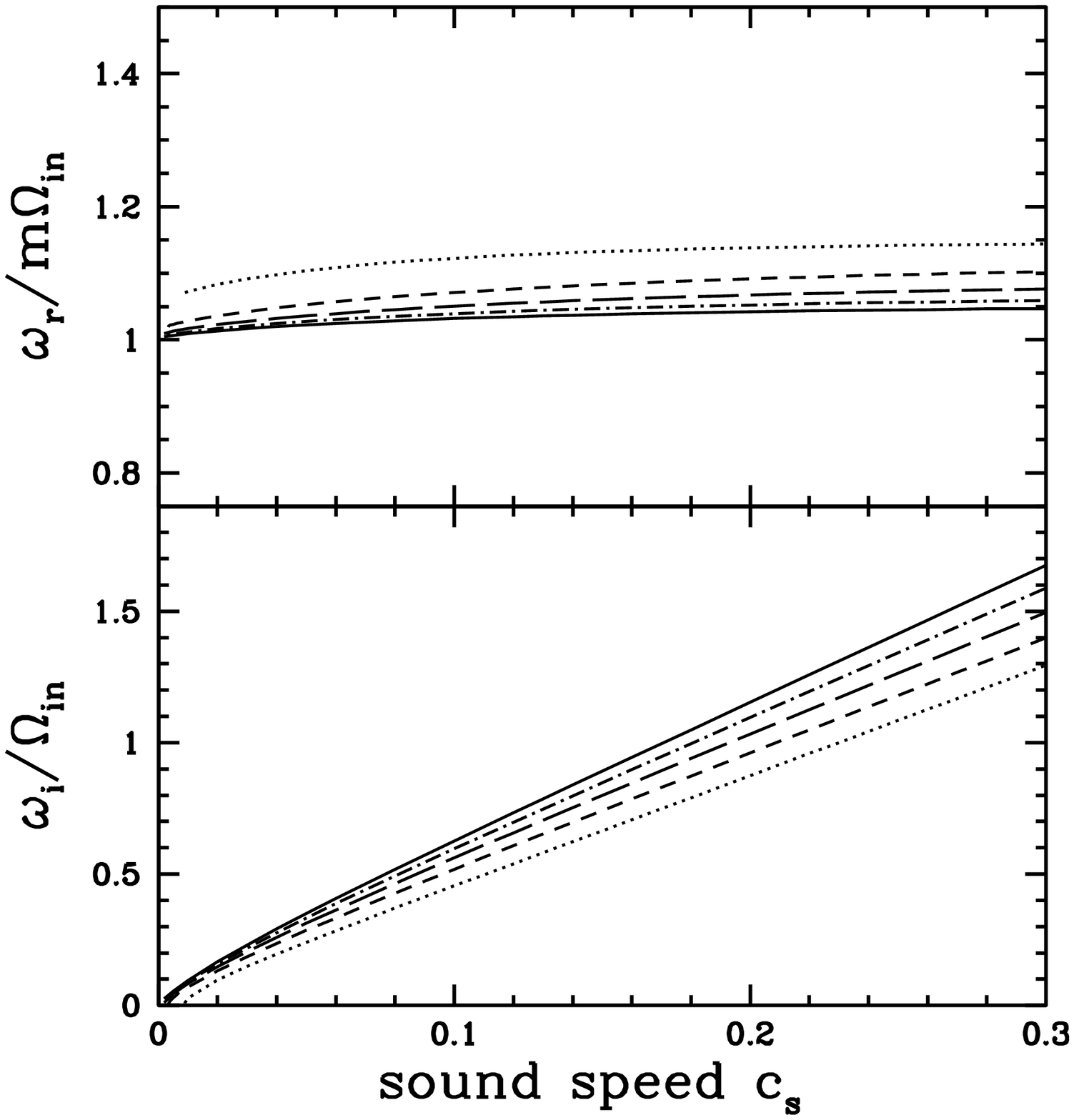,width=0.5\linewidth,clip=} &
\epsfig{file=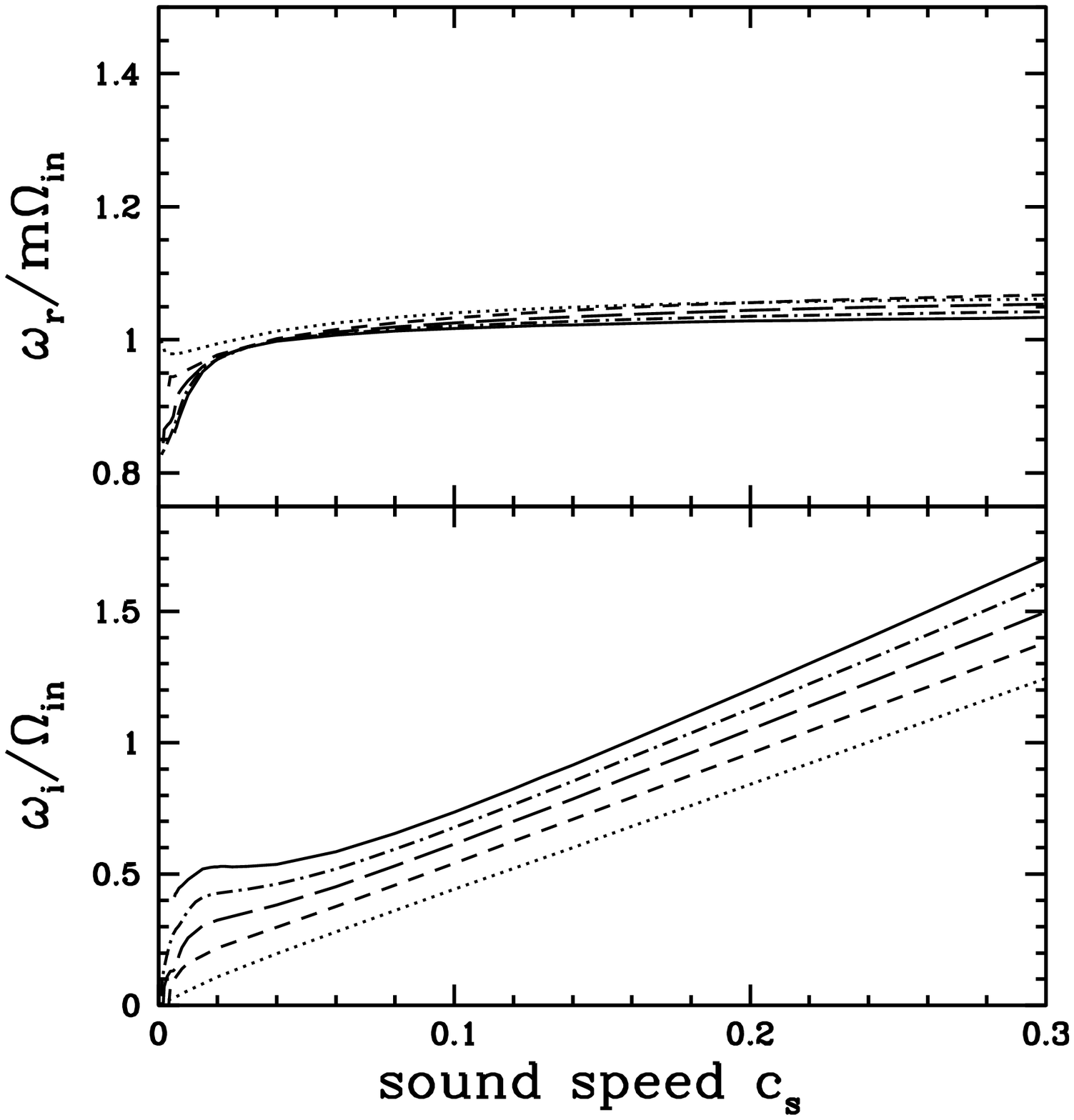,width=0.5\linewidth,clip=}
\end{tabular}
\caption{Real and imaginary frequencies for the interface modes of various $m$ (denoted as in Figure 2) as a function of sound speed $c_s$ with $r_{\rm in} = r_{\rm ISCO}$ for the pseudo-newtonian GR potential. There is no sound speed cutoff for unstable modes as the vorticity is zero at the interface. The left panels show the eigenvalues for $\Sigma_- = 0$ and $\Omega_- = \Omega(r_{\rm in+})$ and the right panel shows the eigenvalues for $\Sigma_- = \frac{1}{99}\Sigma_+$ and $\Omega_- = 0.5 \Omega_{\rm in}$}
\end{figure}

While in Section 3.2 and other sections of the paper we focus on Newtonian discs, it is of interest to consider how general relativity may modify our results. The effect of general relativity can be approximated by using the pseudo-Newtonian Paczynski \& Wiita (1980) potential:
\be
\Phi = -{GM\over r-r_S},
\ee
with $r_S = 2GM/c^2$ the Schwarzschild radius. This gives the Keplerian orbital frequency  ($\Omega_k$) and epicyclic frequency ($\kappa$) as:
\be
\Omega_k = \left(\frac{1}{r}\frac{d\Phi}{dr}\right)^{1/2} =
\sqrt{\frac{GM}{r}}\frac{1}{r-r_S}~, \qquad \kappa =
\left[\frac{2\Omega_k}{r} \frac{d}{dr}(r^2\Omega_k)\right]^{1/2} = \Omega_k
\sqrt{\frac{r-3r_S}{r-r_S}}.
\label{eq:OmegaK}
\ee
with $\kappa \rightarrow 0$ at $r_{\rm ISCO} = 3r_S = 6GM/c^2$.

For $r_{\rm in} \gg r_{\rm ISCO}$ the interface modes are the same as for the Newtonian case. However, as $r_{\rm in} \rightarrow r_{\rm ISCO}$ the suppression effect of $\omega_{\rm vort}^2$ [see Eq. (\ref{simpeq})] is reduced as the vorticity goes to zero at $r_{\rm ISCO}$, so that for $r_{\rm in} = r_{\rm ISCO}$ there is no cutoff sound speed (see Figures 2-3) for interface mode instability. This is illustrated in Figure 4.  Thus if the magnetosphere boundary is at $r_{\rm ISCO}$, the interface modes will always be present and highly unstable for any sound speed.

\subsection{P-modes with Magnetosphere Boundary}
\begin{figure}
\centering
\epsfig{file=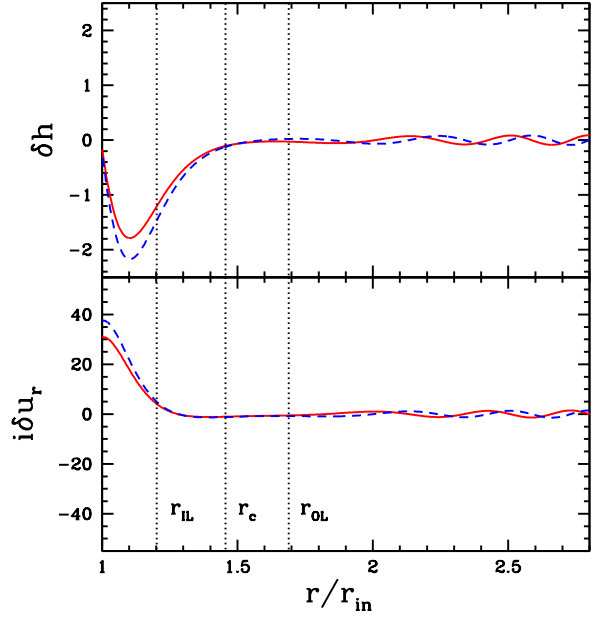,width=0.5\linewidth,clip=}
\caption{The p-mode wavefunctions for $m=4$, $c_s= 0.1 r \Omega$, $\Sigma_-/\Sigma_+ = 1/99$, $\Omega_-/\Omega_+ = 1$ with $\omega/\Omega_{\rm in}=1.914 + 0.000127i$. The real components are shown in solid lines, while the imaginary components are dashed lines. For p-modes the inner and outer Lindblad resonances and the corotation resonance are outside $r_{\rm in}$, denoted by the dotted lines.}
\end{figure}

The boundary condition given by equation (\ref{eq29}) also provides an inner reflection boundary for disc p-modes, which were studied in detail in Lai \& Tsang (2008). These modes  have wavefunctions primarily ``trapped'' in the wave region between the disc boundary $r_{\rm in}$ and the inner Lindblad resonance radius, $r_{\rm ILR}$, where $\omega - m\Omega = -\kappa$. Figure 5 depicts an example of the p-mode wave function for the same disc model as in Figure 1. The growth rates of these p-modes are determined primarily by the outgoing flux at the outer boundary and the effect of the corotation resonance, as discussed in Lai \& Tsang (2008). In Figure 6 the eigenfrequencies are shown for p-modes in a disc with the density profile $\Sigma \propto r^{-p}$, where $p = 3/2$ so that wave absorption at the corotation resonance is inactive (since in this case the vortensity $\kappa^2/(2\Omega\Sigma)$ is constant). For the density profile $p < 3/2$, the corotation absorption tends to damp the mode, while for $p > 3/2$ the corotation absorption enhances it. 

\begin{figure}
\centering
\epsfig{file=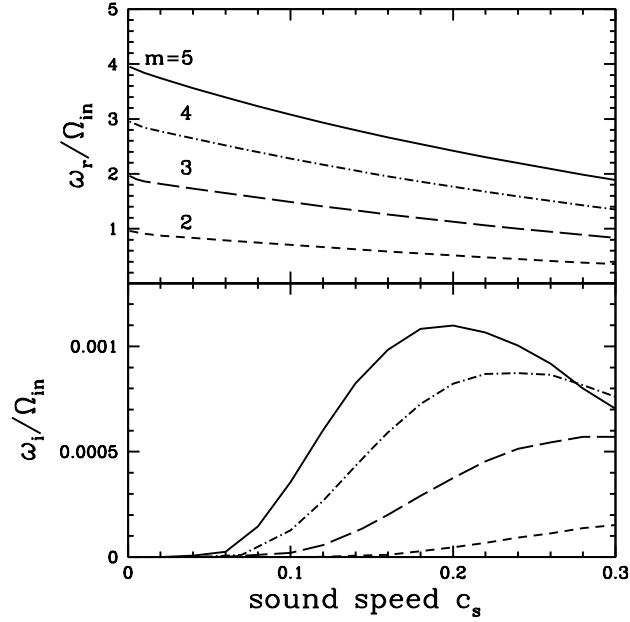,width=0.5\linewidth,clip=}
\caption{Real and imaginary frequencies for p-modes for the magnetosphere boundary condition with $\Sigma_- = (1/99)\Sigma_+$ and $\Omega_- = 0.5 \Omega_{\rm in}$. For $\Sigma_- \ll \Sigma_+$ the p-mode frequencies have very little dependence on $\Omega_-$. Here the disc surface density profile is chosen to be $\Sigma \propto r^{-3/2}$, so that the corotation absorption is inactive and mode growth is purely due to propagation outward at $r_{\rm out}$.}
\end{figure}

\section{Interface Modes at the Star-Disc Boundary}
\subsection{Star-Disc Boundary Condition}
In the case of accretion on to a non-magnetic star, our model consists of a dense uniformly rotating compressible stellar atmosphere truncating the accretion disc. This model ignores the structure of the boundary layer. However the qualitative properties of the dynamics should be captured for modes with characteristic radial length scale much greater than the radial scale of the boundary layer.

Several studies of CVs (e.g. by examining the rotationally broadened line emissions from the stellar surface) have shown that the stellar rotation rates are significantly below the breakup rotation rate (see Warner, 2004), and we limit our examinations to systems with $|\Omega_-| \leq 0.5 \Omega(r_{\rm in+})$.
As in  the case of the disc, we consider only the effect of perturbations on a cylindrical equatorial surface of the stellar atmosphere (i.e., we are considering a "cylindrical" star).
In this region equation (\ref{perturbeq1}) also describes the enthalpy perturbations within the stellar atmosphere. Inside the atmosphere, we assume a small constant density scale height, $-\Sigma/\Sigma' \equiv H_\Sigma \ll r$. Equation (\ref{perturbeq1}) then becomes
\be
\delta h'' - \frac{1}{H_\Sigma} \delta h' - \left(\frac{2m\Omega}{r\tomega H_\Sigma} + \frac{D}{c_s^2}\right) \delta h \approx 0.
\ee
For $H_\Sigma \ll r$ and $H_\Sigma \ll c_s/\Omega$, this has the solution
\be
\delta h \propto \exp[(r-r_{\rm in})/H_\Sigma].
\ee
The Lagrangian pressure perturbation at the stellar surface is then
\ba
\Delta P_- &=& \Sigma_- \left[ \frac{\tomega \delta h}{i\delta u_r} - r(\Omega_k^2 - \Omega^2) \right]_{r_{\rm in-}}\frac{i\delta u_{r}}{\tomega}\nonumber\\
&=& \left[\frac{\kappa^2 - \tomega^2}{\frac{2\Omega m}{r\tomega} - \frac{1}{H_\Sigma}} - r(\Omega_k^2 - \Omega^2)\right]_{r_{\rm in-}} \Sigma_- \frac{i\delta u_r}{\tomega}.
\ea
Once again matching the Lagrangian displacement and pressure perturbation at the interface gives the boundary condition for the interface modes for the star-disc boundary case:
\be
\Sigma_+ \left[ \frac{\tomega \delta h}{i\delta u_r} - \frac{pc_s^2}{r}\right]_{r_{\rm in+}} = \Sigma_- \left[ \frac{\kappa^2 - \tomega^2}{\frac{2\Omega m}{r\tomega} - \frac{1}{H_\Sigma}} + r(\Omega^2 - \Omega_k^2)\right]_{r_{\rm in-}}.\label{starbound}
\ee

\subsection{Numerical Results}
We repeat the numerical procedure of Section 3 using the radiative outer boundary condition [equation (\ref{radout})] and using equation (\ref{starbound}) as the inner disc boundary condition. A sample wavefunction for the star-disc interface mode is shown in Figure 7. For typical mode frequencies the region just outside the boundary is an evanescent zone; wave propagation becomes possible only beyond the outer Lindblad resonance ($r_{\rm OL}$). Figure 8 shows the eigenfrequencies for the lowest order modes with $m = 1,2,\ldots,5$ as a function of disc sound speed, for representative parameters $H_\Sigma = 0.01r_{\rm in}$, $\Sigma_- = 10\Sigma_+$ and $\Omega_- = 0.1 \Omega_+$. Figure 9 shows the dependence of the mode eigenfrequencies on the density ($\Sigma_-$), rotation rate ($\Omega_-$), and scale height ($H_\Sigma$) of the star. We see that both the real mode frequency $\omega_r$ and the growth rate $\omega_i$ do not depend strongly on these parameters.

\begin{figure}
\centering
\epsfig{file=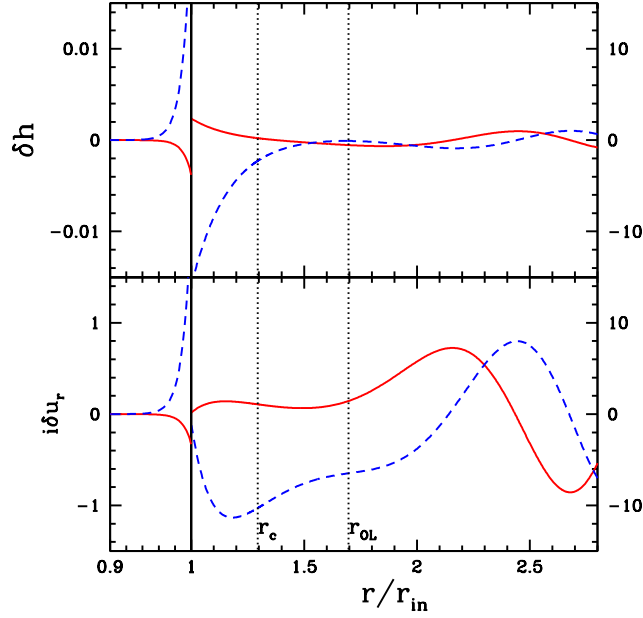, width=0.5\linewidth}
\caption{An example wavefunction for a disc/non-magnetized star interface with $m = 2$, $\Sigma_- = 10 \Sigma_+$, $\Omega_- = 0.1 \Omega_{\rm in}$ and $H_\Sigma = 0.01 r_{\rm in}$ with the eigenfrequency $\omega/\Omega_{\rm in} = 1.378 + 0.0030i$. The left side ($r < r_{\rm in}$) of the plot denotes the perturbation eigenfunctions inside the star, while the right side shows the disc perturbations. The vertical dotted lines denote the corotation resonance radius ($r_c$) and the outer Lindblad resonance radius ($r_{\rm OL}$).}
\end{figure}

\begin{figure}
\centering
\begin{tabular}{ccc}
\epsfig{file=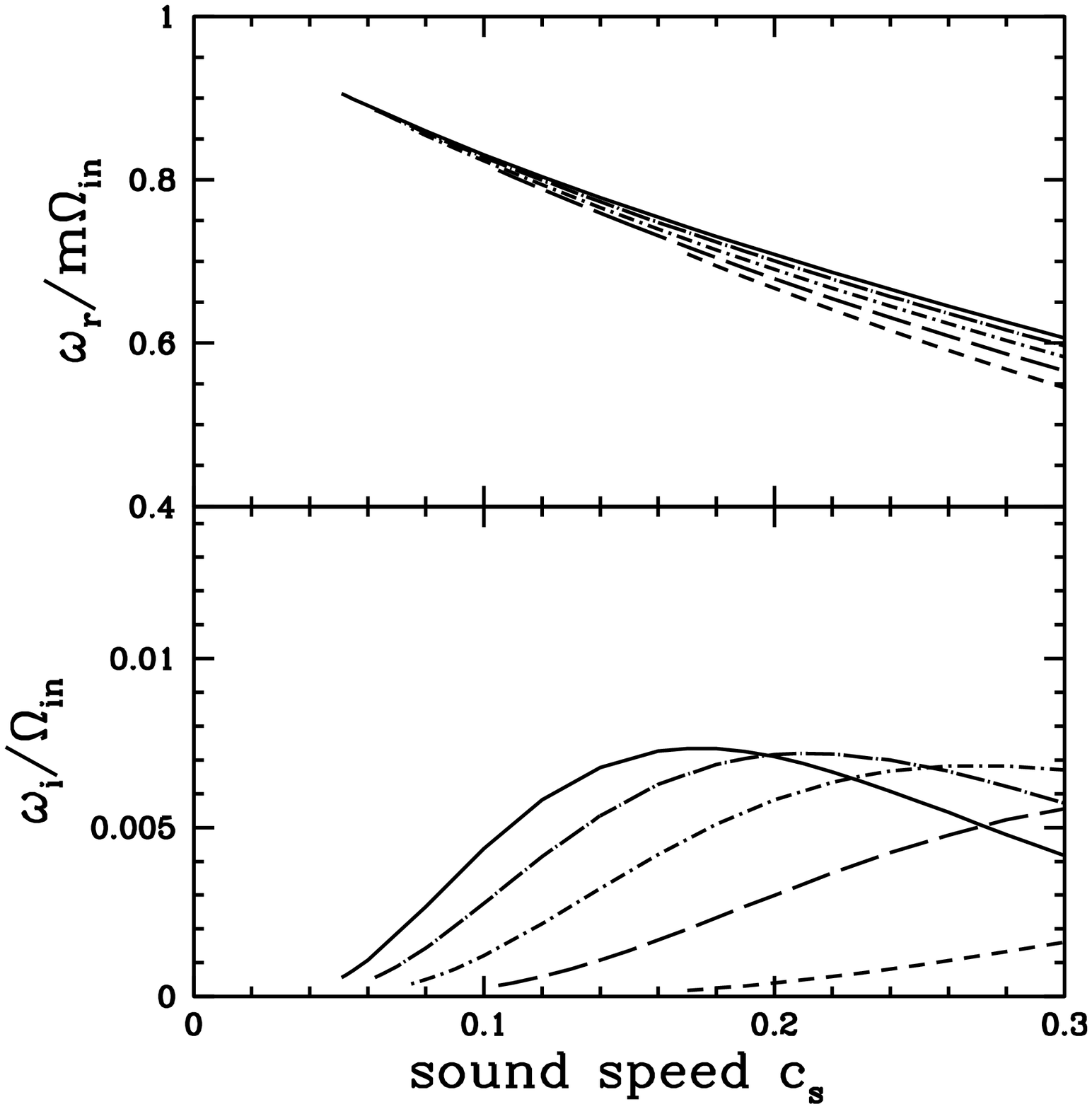,width=0.5\linewidth,clip=} &
\epsfig{file=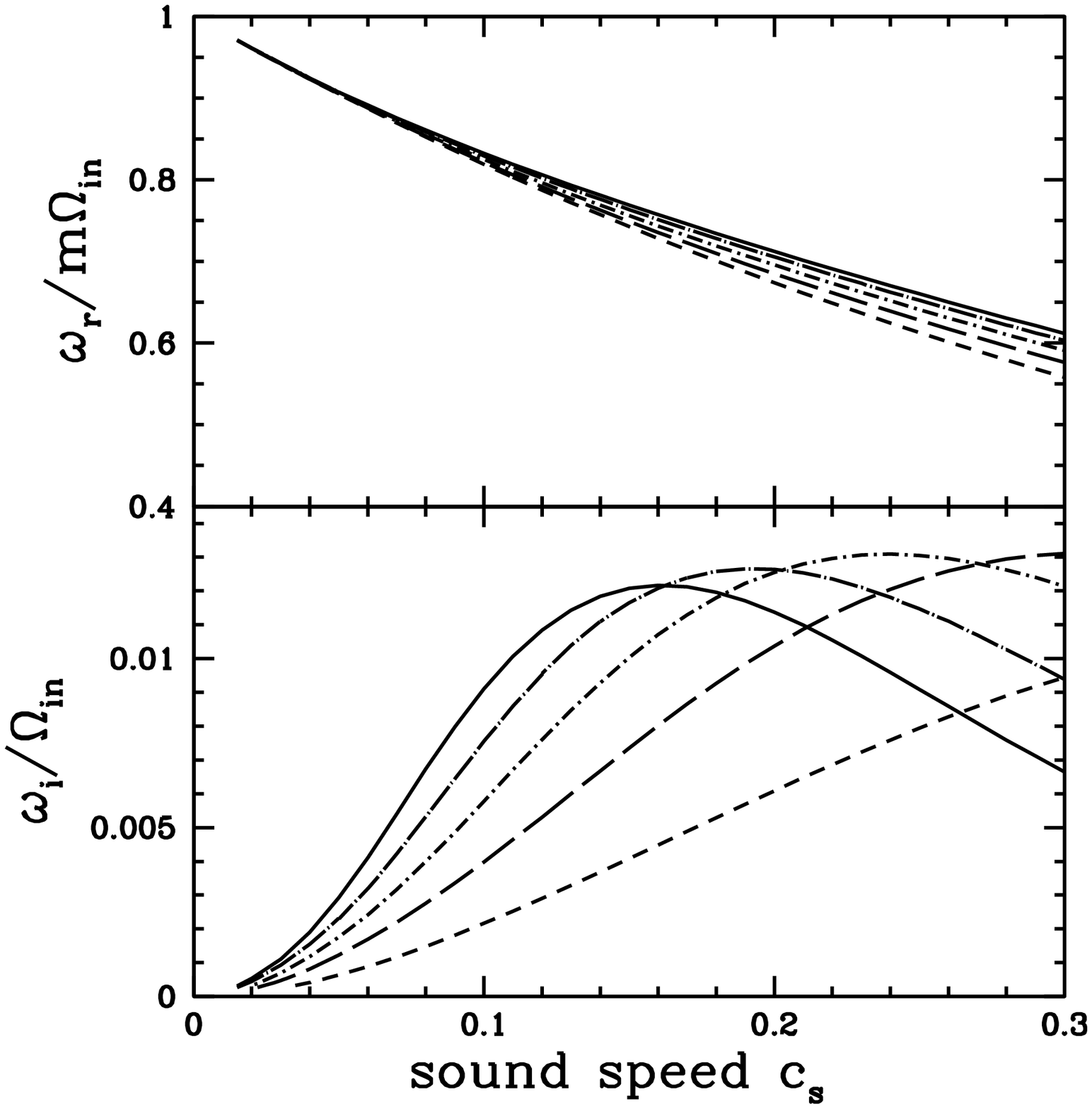,width=0.5\linewidth,clip=}
\end{tabular}
\caption{The eigenfrequencies of the interface modes for the star-disc boundary for $H_\Sigma=0.01r_{\rm in}$, $\Sigma_- = 10\Sigma_+$ and $\Omega_- = 0.1 \Omega_{\rm in}$, as a function of disc sound speed, for $m = 1\ldots 5$. The vertically integrated surface density of the disc is given by  $\Sigma_+ \propto r^{-p}$ with $p = 3/2$ for the left panels, so that corotation absorption plays no role, and $p = 2$ for the right panels, where corotation absorption acts to enhance the mode growth.}
\end{figure}

\subsection{Discussion of Numerical Results}

When the disc is truncated by the star's surface, the effective
gravity acts to stabilize the perturbations (since $\Sigma_- >
\Sigma_+$), as does the vorticity. Thus compared to the interface mode
in the magnetosphere-disc case (Sections 3.1 -- 3.2), the mode growth
rates here are much smaller and are primarily driven by wave
propagation through the corotation, beyond the outer Lindblad
resonance. In the left panels of Figure 8 the eigenfrequencies are
shown for typical parameters [$\Sigma_- = 10 \Sigma_+(r_{\rm in})$,
$\Omega_- = 0.1 \Omega(r_{\rm in+})$, $H_\Sigma = 0.01r_{\rm in}$], and
disc density index $p = 3/2$, so that the corotation absorption plays
no role. For other density indices, wave absorption at the corotation
can act to either damp or grow the interface modes (Tsang \& Lai
2008; Lai \& Tsang 2008). For example, in the Shakura-Sunyaev
$\alpha$-disc model the disc solution for the outer disc solution
(with free-free opacity and gas pressure dominating) has the surface
density $\Sigma \propto r^{-3/4}$, hence the modes would be stabilized
by absorption at the corotation resonance. However, for models where
the disc has density index $p > 3/2$ at corotation, the corotational
absorption acts to enhance mode growth, as shown in the right panels
of Figure 8.

The mode eigenfrequencies have very little dependence on the
properties of the stellar atmosphere ($\Sigma_-, \Omega_-, H_\Sigma$),
as shown in Figure 9. The mode frequencies instead primarily depend on
the disc sound speed, which in turn depends on the accretion
rate. Observations of CVs indicate that DNOs are usually only detected
in high $\dot{M}$ states, with the oscillation period decreasing with
increasing luminosity (Warner 2004). The downward trend of the
$\omega_r/m\Omega_{\rm in}$ as a function of $c_s$ in Figure 8 would
appear to contradict the observed period-luminosity
anti-correlation. But note that in our model, the interface mode
frequency depends on the sound speed at the inner-most disc region and
boundary layer, and it will be necessary to model the thermodynamic
and radiative properties of the boundary layer in order to compare
with observation directly. Also, the oscillations of the type
considered here would yield periods shorter than the surface Keplerian
period, except for the $m=1$ mode. Though higher-$m$ modes would be
more difficult to observe due to the averaging out of the luminosity
variation, most observed DNOs, even those with 1:2:3 harmonic
structure (Warner \& Woudt, 2005) occur with period roughly at or
greater than the corresponding surface Keplerian period. 
These long-period oscillations cannot be explained by the model
considered here.

\begin{figure}
\centering
\epsfig{file=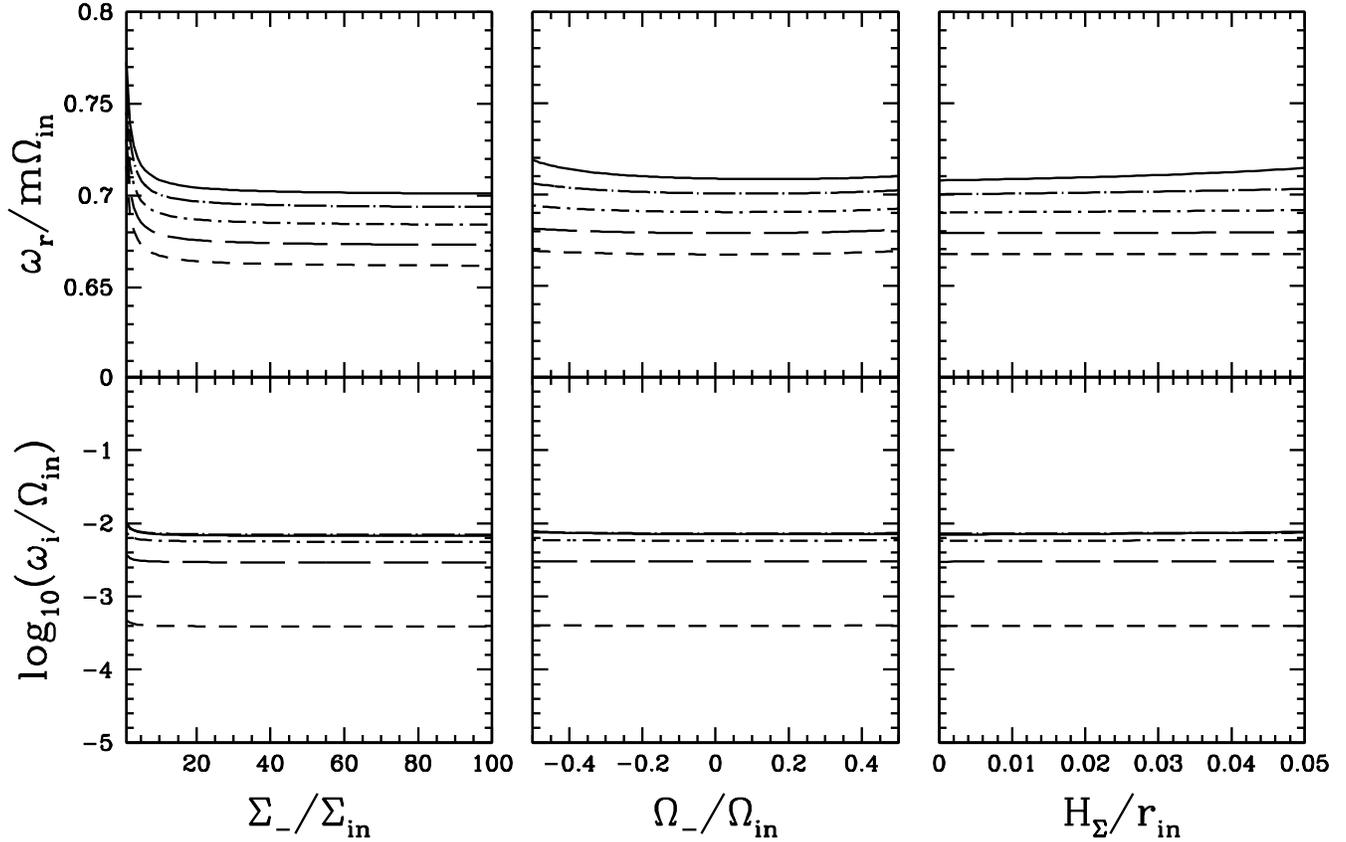, width=\linewidth}
\vskip -3cm
\caption{Eigenfrequencies for disc-star interface modes as a function
of star density ($\Sigma_-$), rotation rate ($\Omega_-$), and
characteristic scale height ($\Sigma_-/\Sigma_-' = H_\Sigma$). The
canonical values for various parameters are: $c_s = 0.2r_{\rm in} \Omega_{\rm
in}$, $\Omega_- = 0.1 \Omega_{\rm in}$, $H_\Sigma = 0.01 r_{\rm in}$
and $\Sigma_- = 10 \Sigma_+$. }
\end{figure}

\section{Conclusions}

We have studied the non-radial oscillation modes at the interface
between an accretion disc and a magnetosphere or stellar
surface. Although the models explored in this paper are perhaps
too simplified compared to realistic situations, they
offer some insight into the behavior of the interface modes in
various astrophysical contexts.

Our study of the interface modes at the magnetosphere-disc
boundary extended the work by Li \& Narayan (2004), who considered
incompressible disc flow (and therefore could not treat real
discs). The model can have very strongly unstable modes due to 
Rayleigh-Taylor and Kelvin-Helmholtz instabilities. In systems where
the magnetosphere has developed from advection of frozen magnetic
flux, the magnetosphere is expected to be roughly rotating with at the
Keplerian rate. Since there is no shear at the interface, only the
Rayleigh-Taylor instability may occur. However the disc vorticty (due
to differential rotation) acts to suppress the instability, leading to
a cutoff below a critical sound speed. Thus a sufficiently hot disc is
required to generate unstable low-$m$ modes. For magnetospheres
rotating with the central star, shearing is expected between the
magnetosphere and disc, and the Kelvin-Helmholtz instability becomes
active. This can help to drive the instability for low-$m$ modes to
overcome the vorticity in low sound-speed discs.  In discs that
terminate near the ISCO in a general relativistic potential, the vorticity
approaches zero at the inner disc radius, and unstable interface modes
can be found for any sound speed.

We can expect a strong dependence of the interface mode growth rates
on the sound speed, and thus accretion rate, while the real
frequencies of these oscillations remain close to $(1-1.3)
m\Omega_{\rm in}$, and do not depend strongly on the sound speed. It
is worth noting that the same boundary condition that gives rise to
the interface modes also gives rise to intertial-acoustic modes (or
p-modes) in the disc.

Although higher-$m$ interface modes are more unstable in our model,
these are less likely to be observed due to the averaging out of the
luminosity variation over the observable emitting region. In addition,
if the effect of viscous damping is considered (e.g., Wang \& Robertson
1985), small wavelength or high-$m$, perturbations are suppressed.

These results are for perturbations with no vertical structure, and
are applicable mainly to the midplane of accretion discs interacting
with a magnetosphere. Global 3D numerical studies of Rayleigh-Taylor
instability induced accretion onto magnetized stars have been performed by
Romanova, Kulkarni \& Lovelace (2008) and Kulkarni \& Romanova (2008),
and show such small $m$ instabilities in the disk midplane. The
low-$m$ oscillations at the magnetosphere-disc interface may be
relevent to the high-frequency QPOs observed in some NS and BH X-ray
binary systems (Li \& Narayan, 2004; see Section 1 of Lai \& Tsang for
a critical review of various theoretical models), although to obtain
the correct QPO frequencies for the BH systems, the disc inner radius
must lie outside the inner-most stable circular orbit.

For the star-disc boundary model considered in in this paper, the interface 
mode growth rates are much smaller than for the magnetospheric case,
since the effective gravity now acts to stabilize the system, and the
Rayleigh-Taylor instability is inactive. The modes discussed here are
unstable due primarily to propagation through the corotation. With
sufficiently steep disc density profile ($\Sigma \propto r^{-p}$ with
$p > 3/2$), corotation absorption can also help to drive these modes,
as studed previously by Tsang \& Lai (2008) and Lai \& Tsang (2008).
Such modes may be responsible for the high-frequency 
(of order the Keplerian frequency at the stellar surface) 
dwarf nova oscillations observed in CVs, although oscillations with longer
periods would require a different explanation.

\section*{Acknowledgements}

This work has been supported in part by NASA Grant NNX07AG81G, NSF
Grant AST 0707628, and by {\it Chandra} Grant TM6-7004X
(Smithsonian Astrophysical Observatory).

\section*{Appendix: Plane Parallel Flow with a Compressible Upper Layer}
Consider a system consisting of two fluids in the gravitational field ${\bf g} = -g \hat{\bf z}$. The upper fluid $(z > 0)$ has density $\rho = \rho_+ e^{-z/H_z}$ (with $H_z = c_s^2/g$, where $c_s$ is the sound speed) and horizontal velocity $u_+$ along the x-axis; the lower fluid $(z < 0)$ is incompressible with density $\rho_-$ and horizontal velocity $u_-$. 

The linear perturbation equations for the upper fluid are
\ba
\frac{\partial}{\partial t}\delta \rho + \nabla \cdot ( \rho \delta {\bf u} + {\bf u} \delta \rho) &=& 0\\
\frac{\partial}{\partial t}\delta {\bf u} + ({\bf u} \cdot \nabla) \delta {\bf u} + (\delta {\bf u} \cdot \nabla) {\bf u}_o &=&  -\nabla \delta h
\ea
where $\delta h = \delta P/\rho$. 
For perturbations of the form $e^{ikx - i \omega t}$, these become
\ba
-i\tomega \delta \rho + i k \rho \delta u_x + \frac{\partial}{\partial z}(\rho \delta u_z) &=& 0\label{momentumpert},\\
i\tomega\delta u_z &=& \frac{\partial}{\partial z}\delta h\label{zmasspert},\\
i \tomega \delta u_x &=& i k \delta h,
\ea
where $\tomega_+ = \omega - ku_+$.
Assuming the perturbation is isothermal, so that $\delta P = c_s^2 \delta \rho$, we obtain
\be
\delta h''(z) - \frac{1}{H_z} \delta h'(z) - (k^2 - \tomega_+^2/c_s^2) \delta h(z) = 0.\label{uppereq}
\ee
The two independent solutions of equation (\ref{uppereq})
\be
\delta h \propto \exp\left[\frac{z}{2H_z}\left(1 \pm \sqrt{1 + 4H_z^2k_z^2}\right)\right],
\ee
where $k_z^2 = (k^2 - \tomega^2/c^2)$. Obviously, the physically relevant solution is 
\be
\delta h \propto \exp(-\tilde{k} z),
\ee
with $\tilde{k} = (\sqrt{1 + 4H_z^2k_z^2} - 1)/2H_z$. This gives, by equation (\ref{zmasspert}) the Eulerian pressure perturbation for the upper region:
\be
\delta P_+ = -\frac{i\tomega_+\rho_+}{\tilde{k}}\delta u_{z+}.
\ee


For the lower region the fluid is an incompressible potential flow with $\delta {\bf u} = {\bf \nabla} \delta \psi$ with $\delta \psi$ satisifying $\nabla^2 \delta \psi = 0$. For z$ < 0$, the appropriate solution is 
\be
\delta \psi \propto \exp(kz).
\ee
This gives $\nabla(i \omega \delta \psi + \delta P_-/\rho_-) = 0$. The Eulerian pressure perturbation in the lower region ($z < 0$) is then:
\be
\delta P_- = \frac{i\rho_-\tomega_-}{k} \delta u_{z-}.
\ee


%

Matching the Lagrangian displacement and Lagrangian pressure perturbation across the boundary between the upper and lower regions we get
\ba
\frac{\tomega_+^2 \rho_+}{\tilde{k}} + \rho_+ g &=& -\frac{\tomega_-^2}{k} + \rho_- g,
\ea
which can be written as the quadratic:
\be
\omega^2 - \left[\frac{\rho_+\frac{k}{\tilde{k}} 2k u_+ + \rho_- 2 k u_-}{\rho_+\frac{k}{\tilde{k}} + \rho_-}\right] \omega + \frac{\rho_+\frac{k}{\tilde{k}} k^2 u_+^2 + \rho_- k^2 u_-^2}{\rho_+\frac{k}{\tilde{k}} + \rho_-} + g\frac{\rho_+ - \rho_-}{\rho_+ \frac{k}{\tilde{k}} + \rho_-}.
\ee
where $\tilde{k}$ has a non trivial dependence on $\omega$. This has a solution
\be
\omega = \frac{k (\tilde{\rho}_+u_+ + \rho_- u_-)}{\tilde{\rho}_+ + \rho_-} \pm \sqrt{-\frac{k^2(u_+ - u_-)\tilde{\rho}_+\rho_-}{(\tilde{\rho}_+ + \rho_-)^2} - \frac{kg(\rho_+ - \rho_-)}{\tilde{\rho}_+ + \rho_-}}
\ee
where $\tilde{\rho}_+ = \rho_+ k/\tilde{k}$.

\end{document}